\documentclass[12pt,a4paper]{article}
\usepackage{amssymb,amsfonts,amsmath}
\usepackage{color}
\usepackage{epsfig, euscript, graphics}

\newcommand{\cut}[1]{}

\newcommand{\ignore}[1]{}

 \newcommand{\beq}{\begin{equation}}
  \newcommand{\eeq}{\end{equation}}
  \newcommand{\beql}[1]{\begin{equation}\label{eq:#1}}

  \newcommand{\beqa}{\begin{eqnarray}}
  \newcommand{\eeqa}{\end{eqnarray}}
  \newcommand{\beqas}{\begin{eqnarray*}}
  \newcommand{\eeqas}{\end{eqnarray*}}
    \newcommand{\bal}{\begin{align}}
  \newcommand{\eal}{\end{align}}
  \newcommand{\bals}{\begin{align*}}
  \newcommand{\eals}{\end{align*}}
  \newtheorem{Theorem}{Theorem}[section]
  \newtheorem{Proposition}[Theorem]{Proposition}
  \newtheorem{Lemma}[Theorem]{Lemma}
  \newtheorem{Corollary}[Theorem]{Corollary}
  \newtheorem{Definition}[Theorem]{Definition}
    
    \newtheorem{Remark}[Theorem]{Remark}

     \newcommand{\bTheorem}{\begin{Theorem}}
\newcommand{\eTheorem}{\end{Theorem}}
\newcommand{\bProposition}{\begin{Proposition}}
\newcommand{\eProposition}{\end{Proposition}}
\newcommand{\bLemma}{\begin{Lemma}}
\newcommand{\eLemma}{\end{Lemma}}
\newcommand{\bCorollary}{\begin{Corollary}}
\newcommand{\eCorollary}{\end{Corollary}}
\newcommand{\bProof }{\begin{proof}}
\newcommand{\eProof}{\end{proof}}
\newcommand{\bRemark}{\begin{Remark}}
\newcommand{\eRemark}{\end{Remark}}	
\newcommand{\bDefinition}{\begin{Definition}}
\newcommand{\eDefinition}{\end{Definition}}

  \newcommand{\Tr}{{\rm Tr}}

  \newcommand{\beqan}{\begin{eqnarray*}}
  \newcommand{\beqar}[1]{\begin{equation}\label{#1}\begin{array}{l}}

  \newcommand{\eeqar}{\end{array}\end{equation}}

\newcommand{\bmat}{\left[\begin{array}{rr}}
\newcommand{\emat}{\end{array}\right]}
\newcommand{\bvec}{\left[\begin{array}{r}}
\newcommand{\evec}{\end{array}\right]}
\newcommand{\btmat}{\left[\begin{array}{rrr}}
\newcommand{\etmat}{\end{array}\right]}

  \newcommand{\cH}{{\mathcal H}}
  \newcommand{\cI}{{\mathcal I}}

  \newcommand{\cP}{{\mathcal P}}

\def\wshop{\vglue0pt\bgroup\color{white}\unitlength=1mm%

\begin{picture}(0,0)(0,60)%
\put(182,15){\makebox(0,0)[]{\bf\small }}%
\put(182,10){\makebox(0,0)[]{\bf\small March}}%
\put(182,5){\makebox(0,0)[]{\bf\small 23}}%
\put(182,0){\makebox(0,0)[]{\bf\small }}%
\end{picture}

\vglue-12mm\egroup}

\color{black} 

\title{Open systems, quantum probability and logic for quantum-like modeling in biology, cognition, and decision making  }

\author{Andrei Khrennikov\\ 
Linnaeus University, International Center for Mathematical Modeling\\  in Physics and Cognitive Sciences
 V\"axj\"o, SE-351 95, Sweden}

\begin{document}
\maketitle

\abstract{The aim of this review is to highlight the possibility to apply the mathematical formalism and methodology of quantum theory 
to model behaviour of complex biosystems, from genomes and proteins to animals, humans, ecological and social systems. Such models are 
known as quantum-like and they should be  distinguished from genuine quantum physical modeling of biological phenomena. One of the  distinguishing  features of quantum-like models is their applicability to macroscopic biosystems, or to be more precise, to information processing in them. Quantum-like modeling has the base in quantum information theory and it can be considered as one of the fruits of the quantum information revolution.   Since any isolated biosystem is dead, modeling of biological as well as mental processes should be based on theory of open systems in its most general form -- theory of open quantum systems. In this review we advertise its applications to biology and cognition, especially theory of quantum instruments and quantum master equation. We mention the possible interpretations of the basic entities of quantum-like models with special interest to QBism is as may be the most useful interpretation.}

\section{Introduction}

The year 2022 was celebrating for quantum information studies -  Aspect, Clauser, and Zeilinger got the Nobel Prize for experimental and theoretical studies on quantum foundations supporting quantum computing, cryptography, and teleportation. This is the good time-point to highlight the not so commonly known output of the quantum information revolution that is often called the second quantum revolution, namely, the  project on applications of the quantum foundations and formalism outside of physics -- {\it quantum-like modeling} \cite{Fest,KHC1} .  

This is review  on {\it quantum-like  modeling} and its applications, with the emphasis of the role of theory of open quantum systems.  Such modeling is built on the methodology and the mathematical apparatus of quantum theory and it is directed to applications to biology, cognition, psychology, decision making, economics and finances, social and political sciences, and artificial intelligence. It is of essential importance to signify that this approach can be explored for macroscopic systems and the system's size is not significant. The quantum-like framework is applicable on all scales, that is to say from proteins and genes to animals, humans, ecological and social systems. The crucial role is played by the character of information processing by a system and matching with the laws of quantum information theory \cite{Jaeger2}. Systems are treated as information processors. Metaphorically one may say that system's ``hardware'', its physical and biological structures, are not so significant, but the system's ``software'' plays the central role. We can speak about {\it quantum bioinformatics} \cite{QIB} which should not be mixed with {\it quantum biophysics} \cite{QBIOP}. The latter studies the genuine quantum physical processes in biosystems, e.g., in cells. 

This review can't reflect all publications on quantum-like modeling. Google search generates 213 000 000 result on ``quantum-like modeling''  in 0,47 sek. (this is really surprising for me). Partially this review represents the content of the coming book \cite{Oquantum} by reflecting related references, especially on applications theory of open quantum systems to cognition and decision making.
The author (in cooperation with Accardi, Asano, Basieva, Ohya, and Tanaka)  was the pioneer in employing quantum information and open systems outside of physics \cite{Accardi}-\cite{AS2017}.  

It is important to point out the immense influence of mathematics in physics, emphasized by many scientists and in particular by E. Wigner \cite{Wigner}. However, mathematical tools commonly used in theoretical and mathematical biology, cognition, and psychology are not so efficient as in theoretical and mathematical physics. Gelfand pointed out to  ``ineffectiveness of mathematics in biology'' \cite{Arnold}. From my point of view the Gelfand's statement has to be reformulated and one would speak about ineffectiveness of mathematics that is {\it commonly used} in biology, cognition and psychology.
I presume someone have an intention to model the micro-systems behaviour, say electrons, atoms, photons, within classical analysis of functions defined on  phase space, $A=A(q,p).$ In this case the one would confront difficulties and soon would notice either the impossibility of such description of quantum phenomena or at least its ineffectiveness. 

Personally I stress effectiveness of the quantum description and do not highlight various no-go statements concerning the impossibility of the classical description 

Physicists explored a new branch of mathematics, theory of operators in complex Hilbert space in order to describe the quantum phenomena in the effective way. And  in quantum physics the noncommutative operator calculus works very well. Similarly, one should search for novel mathematics which is proper for biological and mental phenomena.  The quantum-like approach advertises the same mathematics that was employed in quantum physics, noncommutative operator calculus in complex Hilbert space. Why is it so attractive for discussed applications?  Personally I was mainly driven by specialties of quantum probability (QP) calculus which matches mental phenomena very well. This point will be discussed latter in very detail. But one can look even at the deeper level.  It is useful to extract the basic problems in mathematical modeling of mental phenomena highlighted by the experts in the field. 

The central objective of this paper is to illuminate some directions in development of quantum-like modeling \cite{Accardi}-\cite{AS2017}, \cite{Aerts}-\cite{QPOL1}. First we introduce the motivation for operating with the  quantum formalism and especially  probability outside of physics. Then we  compare classical and quantum probability (CP and QP) theories and set forth the principles of quantum-like modeling of decision making. In particular, consideration of a special quantum-like model, ``decision making via decoherence'' \cite{AS2}, leads to coupling with 
theory of open quantum systems \cite{ING}. The latter is discussed in more detail with highlighting its applications to behaviour of complex biosystems \cite{BIO} (especially the problem of stability), cognition, brain’s functioning, theories of consciousness, emotional coloring of 
perceptions. 

Theory of open quantum systems accommodates the formalism of quantum instruments \cite{DL}-\cite{Oz1}. This formalism realizes the most general quantum state updates. We stress that the combinations of the basic psychological effects such as the question order effect (QOE) and 
 the response replicability effect (RRE), conduct us to inquire the possibility to proceed using the standard quantum measurement theory (with representation of observables by Hermitian operators and the quantum state update via the projection postulate) 
\cite{PLOS}. We demonstrate that this psychological effects combination can be modeled with quantum instruments \cite{ENTROPY,OJMP}. 

This article is written schematically with the minimal introduction into the quantum methodology and mathematical apparatus (see appendix). 

\section{Exploring quantum formalism and methodology}

For newcomers to the field of quantum-like modeling searching for its motivation, I can recommend two handbooks \cite{handbook,Ashby}
(on quantum models in social science and mathematical psychology), especially article \cite{Why} in the first one  and the preface of the second one.  In article \cite{Why} I argue that functioning of biosystems should including cognition and in particular unconscious-conscious interaction be modeled within open quantum systems theory 
(see also \cite{KHRfrontiers}).

\subsection{Edgar Allan Poe's reasoning in favor of quantum-like modeling }

The preface  to the handbook on mathematical psychology \cite{Ashby} is started with the brilliant citation from a story written by Edgar Allan Poe (1845)  entitled ``The Purloined Letter''. In this story a protagonist, Mr. C. Auguste Dupin discussed limits of mathematics applicability:

\medskip

 {\it ``Mathematical axioms are not axioms of general truth. What is true of relation — of form and quantity — is often grossly false in regard to morals, for example. In this latter science it is very usually untrue that the aggregated parts are equal to the whole. [...] two motives, each of a given value, have not, necessarily, a value when united, equal to the sum of their values apart.''}

\medskip

One can be surprised by Poe's doubts in applicability of mathematics (of 19s) century to moral phenomena (cf. with attempts of say Freud to proceed with ``classical mathematics''). He also expressed doubts in validity of the value-additivity law. This is very deep statement and in quantum mathematics it is formulated as  ``eigenvalues of the sum of operators $C=A+B$ are not equal to the sums of the eigenvalues of the summands'', i.e., generally
$$
c_i \not= a_i + b_i.
$$ 
In fact, the violation of the value-additivity law is the key point of von Neumann's no-go theorem \cite{VN}; the first statement on impossibility of classical reduction of quantum theory.  Then, the authors of \cite{Ashby} also pointed out to the noncommutativity effect in conjunctions, $$A \& B \not= B\&A.$$ This order effect is also naturally  formalized in the quantum framework.  In fact, these two effects, the value-nonadditivity and the order ones, are closely connected. In probabilistic terms they jointly expressed in the violation of the formula of total probability, interference of probabilities \cite{INT0,KHR1,KHR5}.
 
The essential part of quantum-like modeling  is devoted to the order effect \cite{Moore}. Its QP-realization in decision making has been done in article \cite{WB} (see also \cite{WSSB14}) .

\subsection{Mathematical models of mental phenomena: Why quantum?} 

We remark that in \cite{Ashby} the discussion is not coupled to quantum-like modeling: the authors  searched for the novel mathematical tools for  psychology, but  their considerations really cry for the appeal to  the quantum formalism. The main message \cite{Ashby} was that a variety of mathematical methods could be explored to solve the problems mentioned by Edgar Allan Poe which I completely agree with.  The quantum formalism should not be treated as pretending to be the unique mathematical tool for modeling of mental phenomena.
A while ago, in response to the developments of using the quantum formalism outside of quantum mechanics, the eminent quantum physicist Anton Zeilinger (the Nobel Prize Laureate 2022)  told me, 

\medskip

{\it ``Why should it be precisely the quantum mechanics formalism? Maybe its generalization would be more adequate for mathemtical modeling  of mental phenomena …''} 

\medskip

And he is right, for the moment, despite its tremendous success, quantum-like modeling is still at the testing stage.
May be one day new, more advanced mathematical formalism will be suggested  for modeling in cognition, psychology, and decision making.

\subsection{Simplicity, elegance, and generality}  

However, from my viewpoint, the quantum formalism is the most successful due to  its simplicity. The reader may be surprised: ``Simplicity? But the quantum theory is the mysterious and very complicated!''  One would immediately recall the famous statement which commonly assigned to  Richard Feynman 

\medskip

{\it ``I think I can safely say that nobody understands quantum mechanics.''} 

\medskip

But here ``understanding'' is related to the interpretation problem of quantum mechanics; its formalism is very simple; it is linear algebra. And in quantum information theory \cite{Jaeger2} which is the most useful for applications, including quantum engineering, linear state spaces are finite dimensional. So, this is the matrix calculus in ${\cal H}= \mathbb C^n.$ Linear evolution is very rapid and this is the advantage of the quantum-like like representation for mental states and corresponding linear processing of them. 

In engineering linear models often appear as approximations of essentially more complex nonlinear ones. Hence, the quantum-like representation of biological and mental phenomena might be just an approximation for more complex nonlinear processes in the living systems. 

We also emphasize the generality of mathematical modeling based on the quantum formalism: the same formalism and methodology cover 
the variety of biological and mental processes. 

\section{Classical vs. Quantum Probability}

This short section  presents the motivation for employing quantum probability (QP), instead of classical probability (CP), in the mathematical modeling in cognitive psychology and  decision making.  We refer to the Kahneman (the Nobel Prize Laureate in economics) and Tversky (the most cited psychologist) works \cite{Ka1}-\cite{Ka5} pointed out that CP using as the basis of decision theory leads to inconsistencies and paradoxes (such as Allais \cite{A} and Ellsberg \cite{Ellsberg} paradoxes; also see Erev et al. \cite{Erev}). Such motivation is not grounded on some foundational  principles. Here we proceed in parallel with quantum physics, created to resolve inconsistencies between classical electrodynamics and experimental data for the black body radiation. 

In cognitive psychology and decision making, the experiments on irrational behavior and probability fallacies generated a lot of statistical data which does not match, at least straightforwardly, with main CP-laws \cite{QL1}. 

To support the use of QP, we highlight the quantum-like paradigm \cite{KHC1} by which context-sensitive systems, including humans, process information in the form of superpositions, i.e., without ambiguity resolution. Such processing can be effectively described by the quantum formalism. Once again, we appeal to effectiveness of the QP description, not to impossibility to explore CP. Generally proving various no-go statements is counterproductive, both in physics and decision making. If one is able to describe physical or mental phenomenon with CP vs. QP and this description is  effective, then one can proceed with CP vs. QP. 

\section{Quantum Formalism for Decision Making}

We recount the fundamentals of the quantum-like  modeling of decision making, e.g., \cite{KHC3,BU0,QL1,QL2}. The basic scheme explores 
the standard quantum measurement theory. The basic components are represented as follows: 
\begin{itemize}
\item questions, problems, and tasks   as quantum observables - Hermitian operators;
\item  belief or mental states of decision makers by quantum states - normalized vectors in a complex Hilbert space or generally density operators; 
\item the quantum state update via the projection postulate.
\end{itemize}
From the very beginning, we highlight that in applications of the quantum measurement theory to cognition and generally biology the crucial role is played by finding the proper formalization of the state update resulting from decision making.  The projection state update is the simplest one and it can't cover all cognitive phenomena. More general state updates are explored in quantum information theory; they are formalized within quantum instruments theory (section \ref{Qinstruments}).

We also mention a special quantum-like model, {\it ``decision making via decoherence''}, coupled with open quantum systems theory (section \ref{decoherence}). 

The problem of the belief-state interpretation is discussed by paying attention to the diversity of possible interpretations 
\cite{KHRWS}.  The latter is one of the disturbing problems of quantum foundations. We recount the basic quantum state interpretations, namely, individual and statistical. We also mention QBism \cite{Fuchs1}-\cite{FSCH5} as may be the most useful framework  for quantum-like  
decision making. 

I have rather strange relation with QBism and its creators. In fact, QBism showed up loudly at the V\"axj\"o
conferences on quantum foundations (since year 2000) \cite{KHR2,Fuchs1,Fuchs2}. I actively struggled against QBism \cite{KHRQB1}, since for me the use of subjectivbe probability in quantum physics and generally in statistical physics was nonsense. In particular, I actively disturbed Christopher Fuchs during his talks in V\"axj\"o by trying to explain him that  probability cannot be assigned to individual physical events.  However, by being more involved in quantum-like modeling for decision making  I started to treat QBism as may be the best interpretation for QP quantum-like decision making \cite{KHRBa,HAVKHRQ1}.

This is the good place to point out to the applications of QBism to decision making in geology within the project on determination of the perspective for intelligence petroleum reservoir characterization, monitoring and management \cite{Qgeology1,Qgeology2}

\section{Quantum and Classical Logic of Thought}
\label{QL7}

We start  this section with the remark that Boole designed classical (Boolean) logic   for ``investigation of the laws of thought'' \cite{Boole2}.

Although I put so much efforts in justification of quantum-like modeling through QP analysis and especially its contextual nature, slowly I started to understand that the seed of cognition's  quantumness (not only of humans, but also other biosystems), is in the logic structure of information processing (see, e.g., articles on quantum-like modeling of the problem of common knowledge and violation of the Aumann theorem \cite{Aumann1,Aumann2} on the impossibility of agreeing to disagree \cite{KAumann1,KAumann2}.

 Quantum logic corresponds to the linear representation of information. The basic law distinguishing classical (Boolean) and quantum logic is the distributivity law, it is violated in quantum logic 
(see article \cite{OJMP1} for the details).  We now briefly recall the basics of quantum logic \cite{BI,Beltrametti} (cf. with classical logic \cite{Boole1,Boole2}).   

The logical operations  are defined on subspaces  of  complex Hilbert space $\cH$ or equivalently on the set of orthogonal projectors $\cP(H).$  Subspaces (projections) are  represents of propositions (events).
Let $P$ be projection and let $L_P$ be its image, $L_P= P H.$  For a subspace $L,$ 
 $P_L$ is the projection on $L.$.  Denote the projection onto the orthogonal complement 
to the subspace $L_P$ by the symbol $\overline{P},$ i.e., $H=L_P \oplus L_{\overline{P}}.$  

Negation of proposition $P$ is represented by $\overline{P}.$   The operations of  conjunction $\wedge$ and  disjunction $\vee$ are defined as follows.   

 Let $P$  and $Q$ be an orthogonal projections representing some propositions. The conjunction-proposition (event) $P \wedge Q$ is defined as the projector on  intersection of subspaces  $L_P$ and $L_Q,$ i.e., $L_{P \wedge Q} =  L_P \cap L_Q.$ 
We remark that this operation is well defined even for noncommuting projectors, i.e., incompatible quantum observables. Moreover, it is commutative:
\begin{equation}
\label{EQ1}
P \wedge Q = Q \wedge P  
\end{equation}
The same can be said about the operation of disjunction. Here subspace $L_{P\vee Q}$ is defined as the subspace generated by the union of subspaces $L_P$ and $L_Q,$ i.e.,    $P\vee Q$ is projector on  this subspace. 
This operation  is also well defined for non-commuting projectors and, moreover, it is commutative:
\begin{equation}
\label{EQ2}
P \vee Q = Q \vee P  
\end{equation}
Thus, quantum logic is commutative logic. This fact is never highlighted. Thus, in quantum reasoning
noncommutativity is not present at the level of the basic operations of quantum logic, conjunction and disjunction.  
In the light of this fact, the following natural question arises:

\medskip

{\it What is the logical meaning of noncommutativity  of quantum operators?}

\medskip

We recall that noncommutativity is commonly considered as the basic mathematical feature of the quantum theory. Hence, it should play 
the crucial role also at the level of quantum logic. The answer is rather unexpected and it is given by Theorem A below. 

As one knows, classical Boolean logic is distributive  \cite{Boole1,Boole2}, i.e., for any three propositions (events) $X,Y, Z,$ e.g., represented by subsets 
of some set, 
 \begin{equation}
\label{EQ2aat}
X\wedge (Y \vee Z)= (X\wedge Y) \vee (X \wedge Z). 
\end{equation}

{\bf Theorem A.} \cite{OJMP1} {\it Let $P,Q,R$ be projections. They are pairwise commutative if and only the distributivity law
(\ref{EQ2aat})  holds for $X,Y,Z= P, Q,R, \bar P, \bar  Q, \bar  R.$} 

Thus, noncommutativity encodes non-distributivity of quantum logic! Hence, the existence of incompatible quantum observables, i.e., represented by noncommuting operators, is equivalent 
to non-transitivity of logical relations between quantum propositions. In particular, Bohr's complementarity principle 
\cite{BR0}-{\cite{Google} reflects this logical structure.  

This statement is especially important for quantum-like modeling of cognition. The existence of incompatible propositions or questions 
is a consequence of non-distributivity of logic used by a quantum reasoner.  

\medskip

{\it Are humans classical or quantum reasoners?} 

\medskip

It depends on the context of information processing. In some situations humans use the distributive Boolean logic, but in other situations they violate distributivity law and quantum logic can be employed to describe mathematically the later form of reasoning.

From the first sight, it seems to be impossible to characterize distributive vs. non-distributive  (classical vs. non-classical) reasoning in experimentally testable way. However, such characterization was obtained in \cite{OJMP1} and it is based on testing of the response replicability effect (RRE). 

The notion of {\it response replicability} plays the important role in quantum physics. This property of observations is also the common feature of human behavior. Suppose that  Alice is asked some question $A$ and she replies, e,g, ``yes''.  If immediately after answering she is asked this question again, then she replies ``yes''  with probability one.  This is    {\it $A-A$ response replicability.} This is the  feature of quantum observables of the projection type. 

Commonly human decision making has another property --   {\it $A-B-A$ response replicability.}  Suppose that after answering the $A$-question with say the ``yes''-answer, Alice is  asked another question $B.$ She replied to it with some answer. 
And then she is asked $A$ again. In the social opinion pools and other natural decision making experiments, Alice definitely repeats her original answer to $A,$ ``yes''. This is $A-B-A$ response replicability.  Combination of $A-A$ with $A-B-A$ and $B-A-B$ response replicability is called {\it the response replicability effect} (RRE).

\medskip

{\bf Theorem B.} \cite{OJMP1} {The projection observables $P$ and $Q$ show RRE  in  a state $\psi$ if and only if the distributive law holds for this state, i.e., 
\begin{equation}
\label{EQ2aa}
[X\wedge (Y \vee Z)] \psi= [(X\wedge Y) \vee (X \wedge Z)] \psi, 
\end{equation}
for $X,Y,Z= P, Q, \bar P, \bar  Q.$}

\medskip

As was shown in article \cite{OJMP1},  RRE can be checked experimentally. For the moment, only one experimental test was done \cite{RREWB} and its design and methodology was questioned by a few researchers, see comments on PLOS One web-page for this article. 
May be article \cite{OJMP1} would attract attention of experimenters to RRE as the test for nonclassicality of hymen logic. 

\subsection{Quantum vs. quantum-like cognition}

We emphasize that quantum-like modeling of cognition should be sharply distinguished from the quantum brain studies  
(see, e.g.,  \cite{RU}-\cite{I})   attempting to reduce information processing by cognitive systems including ``generation of consciousness'' to the quantum physical effects in the brain. However, we do not criticize the quantum brain project, although its difficulties  are well known: e.g.,  the brain is too hot and big, the scales of neuron's operating are too rough comparing with the quantum physical scales.

In quantum-like modeling, it is simply not important whether the genuine quantum physical processes in brain's cells contribute to cognition or not. Generally quantum-like modeling is performed on the meta-level of cognition; it does not concern the biophysical processes in neurons. In this framework a biosystem, in particular, the brain is treated as a {\it black box}  which information processing cannot be described by the {\it classical probability} \cite{K}  (CP) and, hence, the classical information theory. Non-classical probability and information theories are on demand. In particular, in decision making exploring of CP leads to various paradoxes which are typically coupled to irrational behavior of humans. My suggestion \cite{KHC1} was to employ quantum probability (QP) and quantum information theories, instead of the classical ones.  Why should specially quantum theory be involved? This is the complex problem.
  
	\subsection{Classical, quantum or more general probability theories?}
	
There exist a plenty of other models different  from CP and QP. The use of QP in, e.g., decision making, was not derived from some basic principles for cognition and psychology. Commonly QP is used pragmatically - to resolve paradoxes and to have a general probabilistic framework applicable to decision making, in all areas of humanities and economics, as well as in biology. There is no a priory reason to hope that QP would cover all problems which arise in decision making. One might find paradoxes even in QP-based decision theory.
May be other probabilistic models different both from CP and QP should be employed.    

Surprisingly, physicists have the same problem. In contrast to relativity theory, QM was not derived from natural 
physical principles (see Zeilinger \cite{ZE} for the discussion on this problem). There is no reason to expect that all experiments in micro-world would match QP-constraints.

In physics, one typically debates CP vs. QP, classical vs. quantum physics. However, one can even test  whether physics of microsystems can violate the QP-laws, i.e., whether electrons and photons can behave exotically even from  the QP-viewpoint. The corresponding test is given by  the {\it  Sorkin equality} \cite{Sorkin0}  for the three slit experiment. This is really surprising that two and three slit experiments have so different probabilistic structures. The three slit experiment  was done by the Weihs' group (Austria). They did not find  deviations from QP, the Sorkin inequality was not violated  \cite{Sorkin1,Sorkin2}}. Similar experiments can be done for decision making by humans \cite{3SL2} by using the theoretical formalism of article \cite{3SL20} - presentation of the Sorkin equality 
in terms of quantum probability theory.

\section{Biosystems as Open Quantum-like Systems}

Any alive biosystem is an open system and to analyze its behavior is reasonable to take advantage of  the open quantum systems theory, whether the biosystems  are acknowledged as information processors and the open quantum systems theory is treated as a part of the quantum information theory. The latter is the most general information theory comprising the classical information theory as a particular case. Thus, the open quantum systems part of quantum-like modeling concerns information processing in complex biosystems interacting with their environments.  From the information viewpoint even a cell or a protein are very complex systems. 

The challenging problem of mathematical formalization of  unconsciousness-consciousness  interrelation can also be handled 
within the open quantum systems theory \cite{KHRfrontiers,BIOEM}. Consciousness plays the role of an apparatus performing  measurements over unconsciousness. 
This formalism matches well to the Higher Order Theory of Consciousness \cite{Paper}-\cite{LeDoux3}. It is applied to model mathematically the emotional coloring of conscious experiences. Such coloring is framed as contextualization. So, the theory of emotions is coupled to such a hot topic in quantum foundations as contextuality.  Finally, we discuss the Bell type experiments \cite{Bell}-\cite{CHSH} for emotional coloring  \cite{BIOEM} (see also \cite{Conte}-\cite{Behti} for such experiments in cognition and decision making). 

\subsection{What is life? }

Treating of biosystems as quantum-like information processors can expound order stability in them, i.e., present the quantum-like formalization of Schr\"odinger's speculations in his notorious book {\it ``What is life?"} \cite{SCHE} - see article \cite{What?}.  
Schr\"odinger stressed that order stability is one of the characteristic features of biosystems. Entropy can be used as a quantitative measure of order. Then he noted that in physical systems, entropy has the tendency to increase (the Second Law of Thermodynamics for isolated classical systems and dissipation in open classical and quantum systems). In contrast, biosystems beat this tendency.  Schr\"odinger asked: ``How?'' Quantum information and open systems theory may give the answer to this 
fundamental question of modern science.

In \cite{What?} the process of biosystem’s adaptation to the surrounding environment is described by the {\it Gorini-Kossakowski-Sudarshan-Lindblad Equation} \cite{BIO}, where the von Neumann and linear quantum entropies are employed as measures of the disorder degree. This equation describes Markovean evolution, so we work with quantum Markov dynamics. Markoveanity posses the strong constraint on the class of the mental state dynamics. The description of information processing in biosystems with quantum non-Markovean dynamics is more promising, but at the same time it is more complicated.

 We highlight the role of  the special class of quantum dynamics  --  generating the camel-like shape for quantum entropies. Camel's hump represents: 
\begin{itemize}
\item a) the entropy increase in the process of the initial adaptation to the environment; 
\item  b) the entropy decrease at the post-adaptation stage of the dynamics. 
\end{itemize}
Our analysis \cite{What?} is based on numerical simulation and to describe such class of quantum dynamics analytically is must-have. 

\subsection{Order-stability in Complex Biosystem in Spite Instability in Subsystems}

Once again theory of open quantum systems is used  (see paper \cite{NOBORU}) in  attempting to bring more clarification on the question ``What is life?''  We consider a complex biosystem $S$ composed of many subsystems, say proteins, or cells, or neural networks in the brain, i.e., $S=(S_i).$ We study the following problem: 

{\it if the composed system $S$ can preserve the ``global order'' in the situation of increase of local disorder and if $S$ can preserve its entropy while some of subsystems  $S_i$ increase their entropies.} 

It is shown that within quantum information theory the answer is positive \cite{NOBORU}. Entanglement of the subsystems states plays the crucial role. In the absence of entanglement, the increase of local disorder generates the increase of disorder  in the compound system $S$ (as in the classical regime).

\subsection{Modeling of Brain Functioning: from electrochemical to quantum information states}

As an application of the open quantum systems theory to cognition, we suggest a quantum-like model of brain's functioning (see \cite{KHRfrontiers,KHBR1,BIOEM}). In this model
the general approach of quantum-like modeling - to start directly with the quantum information representation of byosystems' states - is broken.  We start with consideration of electrochemical states of neurons encoded in action potentials. Such states generate brain's mental states which are processed with the open quantum systems dynamics. 

The model does not refer to the genuine quantum physical processes in the brain. Hence, it does not suffer of the well known problems of matching of the quantum and neural scales, temporal, spatial, temperature (cf. \cite{RU}-\cite{I})). In this model, uncertainty generated by the action potential 
of a neuron is represented as quantum-like superposition of the basic mental states corresponding to some neural code, e.g.,
quiescent/firing neural code.   

Mathematically neuron's state space  
is described as complex Hilbert space of quantum information states. The state of a neural network is presented in the tensor product of single-neuron state spaces. 

Brain's mental functions perform self-measurements by extracting 
from the quantum information states concrete answers to questions. This extraction is modeled
in the framework of theory of open quantum systems. 

In this way, it is possible to proceed without appealing to state's collapse.
The dynamics of the state of mental function $F$ is described by the quantum dynamical equation. Its stationary states represent 
classical statistical mixtures of possible outputs of $F$ (decisions). Thus, through interaction with electrochemical environment, 
$F$ (considered as an open system) resolves uncertainty that was originally encoded in superposition representing action potentials of neurons. 

\subsection{Decision Making via Decoherence}
\label{decoherence}

The above scheme of resolution of uncertainty through interaction with environment is known as the quantum dynamical  decision making or {\it decision via decoherence} \cite{AS2} - \cite{AS2017}; see also recent article  \cite{eye} on the  experimental study on eye tracking in the process of decision making and its modeling with Gorini-Kossakowski-Sudarshan-Lindblad Equation \cite{BIO}. Here the experimentally observed stabilization in eye tracking matches perfectly with stabilization of the solution of this equation. One of specialties of this model is employing three dimensional space of mental states. The equation is phenomenological, i.e., it is not derived from neurophysiology beyond eye moving  in the process of decision making. We hope that this study may stimulate further cooperation for finding the physical and physiological signatures of the mental state stabilization in the process of decision making.
It is also interesting to perform new experiments with the design similar to the experiment in \cite{eye} and check whether the same phenomenological equation as in \cite{eye} can be in use. 

The most promising realization of this scheme is the ``differentiation'' model \cite{AS3} by which  the mental state experiences step by step state transitions under the influence of surrounding electrochemical environmental factors. The differentiation leads to stabilization of biosystem's state. This model is applicable on all biological and social scales, from cells to ecological and social systems. 

Decoherence is a deep foundational notion. Heuristically it can be interpreted as the loss of quantumness, transition from QP to CP, 
washing out of interference of probabilities (see, e.g., Zurek \cite{Zurek}). It is quantified with linear entropy, the measure of state's purity.  In quantum information theory, typically decoherence is considered as a negative factor disturbing information processing. In the quantum dynamical decision making decoherence plays the constructive role as decisions' generator.

\subsection{Emotional Coloring of Conscious Experiences}

The open quantum systems theory is also used for  mathematical formalization of the consciousness-unconsciousness interaction, the information exchange between them. Consciousness plays the role of a measurement device, it performs observations 
 over the states of unconsciousness. These observations can be interpreted as brain's self-observations. So, human's thoughts and decisions  are generated in the complex process of interaction between unconscious and conscious states. From the viewpoint of quantum 
foundations, we use Bohr's interpretation of the outcomes of quantum measurements as generated in the complex process of interaction 
between a system and measurement apparatus  \cite{BR0}-{\cite{Google}. In particular, these outcomes are not objective properties of a system which could be associated with it before measurement. In the same way the mind is not objective. Bohr's ideology structured within 
open quantum systems theory matches the Higher Order Theory of Consciousness \cite{Paper}-\cite{LeDoux3}.  

The unconscious-conscious framework can be explored for quantum-like modeling of interconnected dynamics of perceptions and emotions \cite{BIOEM}. More generally this framework describes the emotional coloring process for the variety of conscious experiences, including decision making. There are considered two classes of observables: perceptions and emotions. 
These observables are represented by Hermitian operators acting in the corresponding unconscious state spaces (or more generally 
by projection valued measures - PVMs). The total unconscious state space is their tensor product. Emotional coloring is formalized within the quantum contextuality formalism: emotional observables determine contexts.  Such contextualization reduces degeneration of spectra for observables representing conscious experiences, for instance perceptions or decision making. 
Quantum measurement theory serves as the basis for mathematical modeling of generation  of conscious experiences as observations over unconscious states. To model self-measurements performed by the brain, brain's functioning is treated as the cooperative work of  two subsystems, unconsciousness  and  consciousness. They correspond to a system and an observer.  Article \cite{BIOEM} is concluded with an experimental test of contextual emotional coloring of conscious experiences (cf. \cite{Conte}-\cite{Behti}), namely, on the violation of  the CHSH inequality -- the special Bell inequality associated with the names of  Clauser, Horne,  Shimony, and  Holt \cite{CHSH}.

\section{Quantum Instruments in Physics,  Psychology, and Decision Making} 
\label{Qinstruments}  
	
One of the specialties of my research during the last years is exploring quantum instruments \cite{DL}-\cite{Oz1} in applications to psychology and decision making - starting with article \cite{KHRfrontiers} devoted to modeling unconscious-conscious interaction. Quantum instruments are the basic tools of the modern theory of quantum measurements and open quantum systems. They describe the probability distributions of measurements' outcomes and the quantum state transformations generated by measurements' feedback. Thus, a quantum instrument describes both probability and its update - via quantum state update.  

Such updates are not reduced to ones based on the projections, i.e., given by the L\"uders projection postulate- More general state space transformations are also needed, both in quantum physics, especially quantum information theory, and in quantum-like modeling, e.g., in decision making and psychology.   

Instruments cause representation of quantum observables by a {\it positive operator valued measures} (POVM)s, generalized quantum observables.  Typically POVMs are considered as the basic entities of the modern theory of quantum measurements, especially in quantum information theory. However, POVMs are just byproducts of quantum instruments. POVM does not determine uniquely  a state transformation coupled to measurement's feedback on system's state.   

In physics quantum instruments and, in particular, POVMs associated with them  were introduced at the advanced stage of quantum theory's development. However,  modeling of cognition and decision making should be based on quantum instruments even for the basic psychological effects; for example, the combination of the question order and response replicability effects \cite{ENTROPY,OJMP}. The von Neumann measurement theory has the restricted domain of applications \cite{PLOS,FOUND}. 

The quantum instrument formalism is derived from open quantum systems theory through the indirect measurement scheme employing the unitary operator realization of the interaction between a system and a measurement apparatus \cite{Oz1}.    
 
\section{Question Order and Response Replicability Effects and QQ-equality}

The question order  effect (QOE) \cite{Moore} is an effect of the  dependence of the sequential joint probability distribution of answers on the questions' order: 
$$
p_{AB}\not=p_{BA}.
$$
We remark that for classical probability: 
$$
p_{AB}(x,y) = P(\omega \in \Omega: A(\omega)= x, B(\omega)=  y) =
$$
$$
P(\omega \in \Omega:  B(\omega), A(\omega)= x)  = p_{BA} (y,x).
$$
So, no QOE. But the experimental statistical data collected in social opinion pools demonstrated QOE \cite{Moore}.
A simple and natural example, the Clinton-Gore opinion pool \cite{Moore}. In this opinion-polling experiment, people were
asked one question at a time, e.g., 
\begin{itemize}
\item $A$ = ``Is Bill Clinton honest and trustworthy?'' 
\item $B$  = ``Is Al Gore honest and trustworthy?''
\end{itemize}
Two sequential probability distributions were calculated on the basis of the experimental statistical data, $p_{AB}$ and $p_{BA}$ (first question $A$ and then question $B$ and vice verse).  

Wang and Busemeyer \cite{WB} modelled QOE  with projective instruments - PVMs, i.e., with state update given by the L\"uders projection postulate (see also \cite{WSSB14}).
However, this strategy was  challenged  in paper \cite{PLOS}. This challenge is related to RRE. 

As was demonstrated in \cite{PLOS,FOUND}, QOE+RRE combination  cannot be modeled by von Neumann observables (with projection state update).  The authors of  article \cite{ENTROPY} have overcome this difficulty  within quantum instrument theory by using the mathematical construction based on the indirect measurement scheme \cite{Oz1}. The unitary operators describing the interaction of a system and an observable (question) are directly written via their actions. In this model, the systems are humans and the observables  are questions asked to them. 

Article \cite{ENTROPY} contains rather long calculations in Hilbert space. They are not complicated, but might be boring for inexperienced reader. One may be satisfied by the statement that, for QOE+RRE, it is possible to construct quantum instruments (by using the indirect measurement scheme). For more experienced reader, the calculations can serve as the basis example of construction of instruments for various combinations of psychological or social effects.

We remark that, in contrast to QOE that attracted a lot of attention in experimental psychology and decision making and was supported by a plenty of statistical data \cite{Moore}, RRE is not supported by experimental studies. 
We can mention just one experiment \cite{RREWB}. We hope that the result and discussion in paper \cite{OJMP1} would stimulate psychologists to perform experiments to check RRE.

Wang and Busemeyer derived the famous QQ-equality (QQE), the amazing (and unexpected)  property of cognitive and social data \
\cite{WSSB14}.
This is  a non-parametric inequality   to which the probabilities 
of an experiment must satisfy in order for a quantum model to exist for them, as follows:
$$p(AyBn)+p(AnBy)-p(ByAn)-p(BnAy)=0,$$
where $A$ and $B$ correspond to questions with two possible 
outcomes `Yes' and `No'.  The joint  probabilities are the probabilities of receiving given answers to questions $A$ and $B$ in the same order as they appear, 
e.g. $P(AyBn)$ means the probability to obtain negative answer to the question $B$ before obtaining affirmative answer to the question $A.$
QQE can be easily derived within the standard von Neumann measurement theory, with Hermitian 
operators and the projection state update. However, von Neumann theory is not powerful enough to describe the combination of QQE with other psychological effects, e.g., QOE + RRE. 

Ozawa and Khrennikov \cite{OJMP}  proved  that the combination QOE+RRE+QQE can be modeled within the theory of quantum instruments.

\section*{Appendix: A few words about quantum formalism}

Let $\cH$ be a complex Hilbert space, for simplicity we recstric consideration to finite dimensional spaces.
We recall that a {\it pure quantum state} can be represented by  a normalized by 1 vector of $\cH,$ i.e., 
$||\psi||=1.$  Two vectors that differ only by a phase, i.e., $\psi= e^{i\theta} \phi,$ represent 
the same quantum state. Under consideration of a single state, its phase does not play any role, but by manipulating 
with a few states, the relative phases play the crucial role, e.g., in the interference effect.

A density operator $\rho$ is determined by the conditions: 
\begin{itemize} 
\item $\rho= \rho^\star$ (so, it is a Hermitian operator) 
\item $\rho \geq 0, $
\item $\rm{Tr} \rho=1.$ 
\end{itemize}
The space of density operators is denoted by the symbol $D\equiv D(\cH).$ 

Density operators represent mixed quantum states, statistical mixtures of pure states.
We note that each pure state $\psi$ can be represented by a density operator, projection on the state vector, $\psi.$ 

The space of linear Hermitian operators in $ H$ is linear space over real numbers. 

We  consider linear operators acting in it, {\it superoperators}. A superoperator is called positive if it maps the set of  positive operators into itself: for $\rho \geq 0, \; T(\rho) \geq 0.$

Any map $x \to \cI_A(x),$ where for each $x,$ the map  $ \cI_A(x)$ is a positive superoperator and 
\begin{equation}
\label{sum}
\sum_x \cI_A(x): D \to D
\end{equation}
is called {\it quantum instrument.} It represents one of measurement procedures of an observable $A.$

 The probability for the output $A=x$ is given by the Born rule in the form 
 \begin{equation}
\label{BRULEy}
P(A =x|\rho) = \Tr\; [\cI_A(x) \rho].
\end{equation}
We note that 

 Measurement with the output $A=x$ generates the state-update by  transformation 
\begin{equation}
\label{TRA4}
\rho \to \rho_x= \frac{\cI_A(x)\rho}{Tr \cI_A(x)\rho}.
\end{equation}

An observable $A$ can be measured  by a variety of instruments generating the same probability 
distribution, but different state updates. 

Let 
$$
\cI_A(x)\rho= P\rho P
$$
where $P$ is a projection. Such instrument is called projection instrument-

\end{document}